\documentclass[12pt,a4paper]{article}
\usepackage[margin=1in]{geometry}
\usepackage{authblk}
\usepackage[utf8]{inputenc}
\usepackage{mathtools}
\usepackage{amsmath,amssymb}
\usepackage{braket}
\usepackage{subfiles}
\usepackage[dvipsnames]{xcolor}
\usepackage{tikz}
\usetikzlibrary{arrows.meta}
\usepackage{graphicx}
\usepackage{enumerate}
\usepackage{tabularx}
\usepackage{subcaption}
\usepackage{caption}
\usepackage[language=english,backend=biber,citestyle=numeric-comp,style=chem-acs,sorting=none,doi=true]{biblatex}
\usepackage[T1]{fontenc}
\usepackage{newpxtext,eulerpx}
\usepackage{bm}
\usepackage[colorlinks]{hyperref}
\DeclareUnicodeCharacter{2212}{-}

\AtBeginDocument{
    \hypersetup{
    linkcolor=cyan,
    citecolor=Cerulean,
    urlcolor=Cerulean,
    allbordercolors=white
    }
}

\title{Unravelling the fluorescence kinetics of light-harvesting proteins with simulated measurements}
\author[1]{Callum Gray} 
\author[2]{Lekshmi Kailas} 
\author[2]{Peter G. Adams}
\author[1,*]{Christopher D. P. Duffy}
\affil[1]{School of Physics and Astronomy, University of Leeds, Leeds, LS2 9JT}
\affil[2]{School of Biological and Chemical Sciences, Queen Mary University of London, Mile End, London E1 4NS}
\affil[*]{c.duffy@qmul.ac.uk}
\begin{document}

\maketitle

\begin{abstract}
    The plant light-harvesting pigment-protein complex LHCII is the major antenna sub-unit of PSII and is generally (though not universally) accepted to play a role in photoprotective energy dissipation under high light conditions, a process known Non-Photochemical Quenching (NPQ). The underlying mechanisms of energy trapping and dissipation within LHCII are still debated. Various proposed models differ considerably in their molecular and kinetic detail, but are often based on different interpretations of very similar transient absorption measurements of isolated complexes. Here we present a simulated measurement of the fluorescence decay kinetics of quenched LHCII aggregates to determine whether this relatively simple measurement can discriminate between different potential NPQ mechanisms. We simulate not just the underlying physics (excitation, energy migration, quenching and singlet-singlet annihilation) but also the signal detection and typical experimental data analysis. Comparing this to a selection of published fluorescence decay kinetics we find that: (1) Different proposed quenching mechanisms produce noticeably different fluorescence kinetics even at low (annihilation free) excitation density, though the degree of difference is dependent on pulse width. (2) Measured decay kinetics are consistent with most LHCII trimers becoming relatively slow excitation quenchers. A small sub-population of very fast quenchers produces kinetics which do not resemble any observed measurement. (3) It is necessary to consider at least two distinct quenching mechanisms in order to accurately reproduce experimental kinetics, supporting the idea that NPQ is not a simple binary switch switch.  
\end{abstract}

\section{Introduction}

The light-harvesting antenna system of Photosystem II (PSII) in higher plants is a large, modular assembly of pigment-protein complexes, the major one being the cyclic heterotrimer LHCII~\cite{wei_structure_2016}. Under low intensity illumination the PSII antenna operates with a quantum efficiency of 0.8 - 0.85~\cite{Bjorkman_FvFm_1987} though this drops significantly at higher intensities~\cite{Maxwell_Johnson_2000}. This reflects the non-photochemical quenching (NPQ) mechanism, a photoprotective response involving the (mostly) reversible down regulation of PSII antenna efficiency following a sudden increase in light levels~\cite{ruban_nonphotochemical_2016,ruban_chlorophyll_2022-2}. The purpose of this is to mitigate \emph{photoinhibition} of PSII~\cite{aro_photoinhibition_1993}, the slowly reversible (and metabolically costly) oxidative damage to the PSII reaction centres.    

Why the exact mechanism of NPQ is still the matter of debate but a general (though not universal) consensus on the basic scheme has emerged in the last decade. The primary trigger is the formation of a steep $\Delta\text{pH}$ across the thylakoid membrane~\cite{horton1996regulation}. This interacts with the PSII antenna, predominantly LHCII trimers but possibly also in minor (monomermic) antenna complexes and even the PSII core~\cite{Nicol_site_NPQ_2019}, resulting in both subtle conformational change in the individual LHCII complexes ~\cite{ruban_identification_2007,liguori_different_2017,daskalakis_conformational_2020} and their mutual clustering/aggregation~\cite{horton1991aggregation,BETTERLE200915255,Johnson2011aggregation,,chmeliov_aggregation-related_2019,Kim2020cluster,Tutkus2021Aggregation}. This somehow alters the interactions between the pigment molecules within LHCII, creating molecular states (known as \emph{quenchers}) that trap and dissipate excitations where before they promoted rapid excitation transfer to the reaction centres. Also involved are the PSII subunit PsbS~\cite{li_pigment-binding_2000,li_psbs-dependent_2002} and the enzymatic de-epoxidation of the LHCII-associated carotenoid violaxanthin to zeaxanthin (the xanthophyll cycle~\cite{Demmig1987xanth}), though both are allosteric regulators of NPQ rather than pre-requisite triggers~\cite{Wilson2020review}.      
    
The molecular details of the quencher states and the structural changes that form them are still unclear, and for a comprehensive discussion of the various proposed models the reader is directed to several reviews~\cite{derks_diverse_2015,ruban_nonphotochemical_2016,Wilson2020review} and a collected work~\cite{DemmigAdams2014NPQ}. Briefly, most models assume that the quenchers involve the various carotenoid pigments bound by the PSII antenna, which are attractive candidates due to their inherently short excited state lifetimes, $ \tau_{\text{car.}} \sim 10\text{ps} $ ~\cite{polivka_ultrafast_2004}, compared to that of chlorophyll (Chl), $\tau_{\text{Chl}}\sim 4 ns$. Some suggest quenching via excitation energy transfer (EET) from a low energy cluster of Chls to the carotenoid lutein within LHCII~\cite{ruban_identification_2007,duffy_modeling_2013,liguori_different_2017,son_flipping_2019,son_observation_2020,son_observation_2020}, while others propose chlorophyll-carotenoid charge transfer (CT)~\cite{park_chlorophyllcarotenoid_2019,cupellini_charge_2020} or \emph{excitonic}~\cite{van_amerongen_understanding_2001,bode_regulation_2009,balevicius_excitation_2020} states. Not all models involve carotenoids, instead proposing excitation quenching Chl-Chl CT states~\cite{muller_singlet_2010,ostroumov_characterization_2020}. These models all differ in the kinetics, location and density of quenchers, and experimentally determining which are involved in \emph{in vivo} NPQ (at least definitively) is difficult.  

Experimental measurements of quenching roughly fall into two categories: ultra-fast time-resolved spectroscopy applied to isolated complexes, or lower-resolution fluoresence techniques applied to both isolated LHCII and more intact systems. The first includes transient absorption (TA) measurements of LHCII aggregates/oligomers in aqueous suspension~\cite{ruban_identification_2007,muller_singlet_2010,van_oort_revisiting_2018}, trimers immobilized in gel~\cite{saccon_protein_2020}, and more recently two-dimensional (2D) electronic spectroscopy measurement on LHCII in lipid nanodiscs~\cite{son_flipping_2019,son_observation_2020,Son2021_nanodiscs}. The measured kinetics have been fit to various kinetic models or qualitatively compared to quantum mechanical simulations of a single LHCII trimer~\cite{duffy_modeling_2013,fox_possible_2018,balevicius_fine_2017,cupellini_charge_2020,lapillo_energy_2020}, but they have yet to reveal a definitive mechanism (or mechanisms). Moreover, induction of quenching in these \textit{in vitro} LHCII systems generally requires low detergent conditions, meaning these quenching processes maybe different to those that occur \emph{in vivo}. The second category, fluoresence techniques, includes Pulse Amplitude Modulation (PAM) fluorescence measurements which were originally used to probe the kinetics of NPQ formation and relaxation in intact plants and membranes~\cite{OxboroughHorton1988} and various fluorescence lifetime measurements of leaves and chloroplasts~\cite{Gentyfluoresence1992,johnson_photoprotective_2009,chmeliov_aggregation-related_2019}, LHCII aggregates in solution~\cite{pascal_molecular_2005,fox_possible_2018}, in liposomes~\cite{natali_light-harvesting_2016,Hancock2019,Tutkus2021Aggregation,manna_membrane-dependent_2021} and on mica surfaces~\cite{adams_correlated_2018}. 

While fluorescence lifetime measurements have a lower time resolution than TA/2D, they do probe a critical part of the NPQ mechanism: the competition between the quenchers and long-range energy transfer between neighbouring LHCII trimers. Analysis of the fluorescence decay kinetics therefore involve course grained models of energy transfer and quenching/trapping within large LHCII assemblies~\cite{valkunas_excitation_2011,chmeliov_nature_2016,chmeliov_aggregation-related_2019}. In 2011, Valkunas \emph{et al.} published such a model, addressing whether fluoresence decay kinetics could discriminate between different proposed quenching mechanisms~\cite{valkunas_excitation_2011}. They assumed that the aggregate was composed of quenched and unquenched LHCII that could exchange energy. They defined three characteristic timescales: the migration time, $\tau_{mig}$, which characterized energy transfer from the antenna bulk to the vicinity of the quencher, the \emph{transfer-to-trap} time, $\tau_{tt}$, for transfer to the quenched site from its nearest neighbour, and the quenching time, $\tau_{Q}$, the actual timescale of excitation dissipation. They then considered two limiting quenching scenarios, $\tau_{tt}>>\tau_{Q}$, termed \emph{slow/fast} (S/F), and $\tau_{tt}<<\tau_{Q}$  (\emph{fast/slow} or F/S). They showed that at low-excitation density (much less than one excitation per LHCII) the fluorescence decay kinetics of the two models were both essentially mono-exponential. However, at higher excitation densities, where quenching must compete with excitation annihilation, the S/F kinetics became bi-exponential, which is consistent with experimental observation. This was judged to support the idea of quenching via EET to a carotenoid over a mechanism such as Chl-Chl CT states, since EET is relatively slow ($10-100$ ps) and dissipation by the Car excited is relatively fast ($< 10$ ps).
. 

Here we revisit the idea of identifying the quenching mechanism from the fluorescence decay kinetics of LHCII aggregates, addressing several limitations in previous models. Firstly, previous models tend to simulate the \emph{excitation} decay kinetics of the aggregate following instantaneous excitation. Here we simulate the actual \emph{fluorescence} decay kinetics as measured by Time-Correlated Single Photon Counting (TCSPC), which is a wide-spread and accessible experimental technique used to (indirectly) measure excitation decay. In TCSPC, sample excitation is via a laser pulse of finite ($\sim 10- 100$ ps) temporal width which will overlap with fast processes such as annihilation. Measured traces are also noisy and analysed by fitting lifetime components which may obscure any fine kinetic differences between different quenching mechanisms. Secondly, thanks to various TA/2D measurements and atomistic models of LHCII, we have more detailed hypotheses about the mechanism of excitation trapping and dissipation. By combining more detailed models of the quenching mechanism with a realistic simulation of a typical measurement we extract a surprising amount of detail from simple fluorescence measurements. In a meta-study of several published fluorescence experiments we show that observations are consistent with most LHCII trimers switching to a relatively weak quenching state and that at least two, kinetically distinct, mechanisms are invovled.

\begin{figure}[ht]
\centering
\includegraphics[width=12cm]{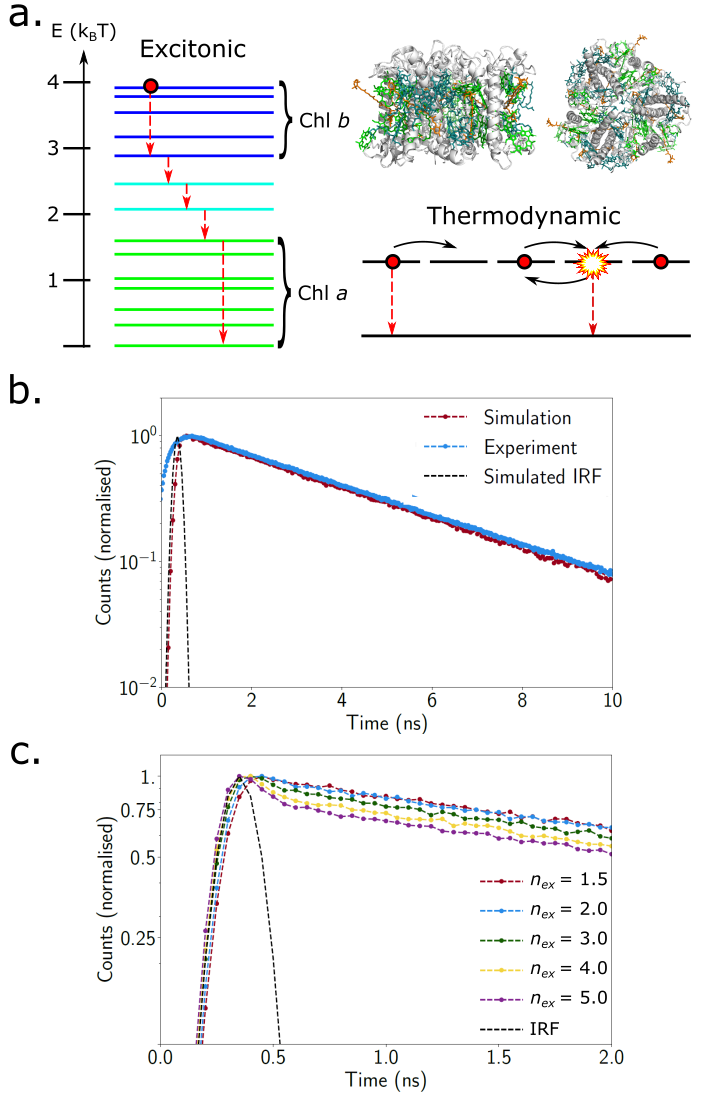}
\caption{\textbf{a.} Diagram of how we model a single LHCII trimer inset, shown side and top view. On the \textbf{left} is a sketch of a typical excitonic model of the complex. A ladder of delocalized exciton states are formed by mixing of the Chl excited states. States that are predominantly Chl \textit{a}/\textit{b} are represented in green/blue, with teal indicating mixed states. Following photo-excitation the exciton (red dot) rapidly relaxes to the Chl a states. On the \textbf{right} is the thermodynamic model where exciton states are replaced with a smaller number of \emph{thermodynamically equivalent} effective states. The excitons can internally sample each of these states. They can either relax to the ground state (via fluorescence or internal conversion) or, if there is more then one exciton, non-radiatively via singlet-singlet annihilation (explosion icon). \textbf{b.} Simulated and experimental TCSPC traces for LHCII trimers isolated in dilute solution in annihilation free conditions. Experimental data is from~\cite{adams_correlated_2018}, while simulated trace assumed $n_{ex}=0.05$. \textbf{c.} Simulated traces at increasing excitation density.}
\label{fig1}
\end{figure}

\section{Theoretical Framework}

Our analysis of TCSPC kinetics of LHCII aggregates is based on a model with two couplled sub-layers.
The first layer is the kinetic Monte-Carlo simulation of photon absorption, energy migration, exciton annihilation, non-radiative decay, fluorescence and quenching within a lattice of coarse-grained LHCII trimers. These are the processes that reflect the internal function of the LHCII assembly but, apart from fluorescence, they are not observable. The second layer deals with the observable, binning of fluorescence decays in the same manner as in a TCSPC experiment and performing multiple re-convolution fits of the resulting decay curves, effectively treating it as if it were experimental data. 

\subsection{Thermodynamic model of an isolated LHCII trimer}

The basic assumption of this work is that the internal excitation equilibration within an LHCII timer are much faster than energy transfer between trimers and deactivation processes like fluorescence, internal conversion and annihilation. In detailed models of the former~\cite{novoderezhkin_energy-transfer_2004}, the coupling between Chl electronic transitions gives rise to a manifold of delocalized exciton states. Following photo-excitation the exciton rapidly ($1-2$ ps) equilibrates across the lowest energy (predominantly Chl \textit{a}) states (see the left side of Fig. \ref{fig1} \textbf{a.}). In our model we replace the $N_{\epsilon}$ exciton states with $\tilde{N}<N_{\epsilon}$ \textit{thermodynamically equivalent} states (see right side of Fig. \ref{fig1} \textbf{a.}), 
\begin{equation}
\tilde{N}=\Biggl\lceil\sum_{i=1}^{N_{\epsilon}}\exp{\left(-\frac{\epsilon_{i}}{k_{B}T}\right)}\Biggr\rceil
\end{equation}
where $\epsilon_{i}$ is energy of the $i^{th}$ exciton state relative to the lowest one, which are taken from~\cite{novoderezhkin_energy-transfer_2004}. At $T=300$ K $\tilde{N}\approx 15$ (5 states per monomer subunit). If $n(t)\leq\tilde{N}$ is the number of excitations within the trimer then the equation of motion for an ensemble of of isolated LHCII trimers is,
\begin{equation}
\frac{d}{dt}n = \sigma J \left(\tilde{N} - n\right) - \sigma_{\text{SE}}J(t)n - k_{\text{decay}} n - \gamma n \left(n - 1\right)
\label{eq:dndt_trimer}
\end{equation}
where $J(t)$ is the photon flux from the excitation pulse, and $\sigma$ and $\sigma_{\text{SE}}$ are the absorption and stimulated emission (SE) cross-sections of the trimer respectively. $k_{decay}$ and $gamma$ are rate constants for excitation decay (fluorescence and internal conversion) and annihilation respectively. Generally, we assume $\sigma \sim 2\times10^{-15} \text{cm}^{2}$, although this varies strongly with excitation wavelength. In Fig. \ref{fig1} \textbf{b.} we show the a TCSPC trace for LHCII in solution with an excitation wavelength of $\lambda = 485$ nm~\cite{adams_correlated_2018}, where absorption by carotenoids will be significant. Here, rather than estimate a particular value for $\sigma$, we define, 
\begin{equation}
n_{ex}=\sigma \int_{-\infty}^{\infty} dt J(t)
\label{eq:nex}
\end{equation}
as the average number of excitations per trimer one would have \emph{if} all of the energy delivered by the pulse arrived simultaneously ($J(t)\rightarrow\delta(t)$). We can therefore define excitation conditions in terms of $n_{ex}$ which is agnostic of any particular experimental conditions. As a rough guide, if the excitation wavelength is $485$ nm and we assume $\sigma \sim 1\times 10^{-15} \text{cm}^{2}$ (see~\cite{barzda_fluorescence_2001}) then $n_{\text{ex}}=0.1$, $1.0$ and $5.0$ would require laser fluences, $\int_{-\infty}^{\infty} dt J(t)\sim 0.04$,  $0.4$, and $2.0$ mJ cm$^{-2}$ respectively. These laser fluences are typical of those used in spectroscopy experiments although annihilation is generally purposely avoided in measurements by keeping to the lower end of this range.     

$\sigma_{\text{SE}}$ is also difficult to define absolutely since complex excitation equilibration within the pulse duration. We assume that $\sigma_{\text{SE}}\sim\sigma$, meaning that SE becomes more likely than absorption when the trimer is half-filled with excitations.   

$k_{\text{decay}}=k_{IC}+k_{F}$ is composed of internal conversion and fluorescence but rather than specify both we realise that $k_{IC}$ effectively rescales $k_{F}$ and $k_{\text{decay}}$ is an effective fluorescence rate. In Fig. \ref{fig1} \textbf{b.} we fit a model with $n_{ex}=1$ and $k_{\text{decay}}^{-1}=3.6$ ns to an experimental TCSPC trace for LHCII in solution, assuming a $200$ ps (FWHM) Gaussian pulse (i.e. the Instrument Response Function, IRF). The fluorescence kinetics are purely mono-exponential. 

For $n_{ex}>1$ annihilation effects become apparent, though typical TCSPC experiments will tune laser fluence to avoid this. The annihilation time constant, $\gamma^{-1}$, in LHCII aggregates has been measured via TA, with values $16-24$ ps depending on the model used to analyse the TA kinetics~\cite{v_barzda_singlet-singlet_2001}. We take $16$ ps throughout this work. Fig. \ref{fig1} \textbf{c.} show how annihilation introduces a second faster component to the TCSPC kinetics of an isolated LHCII trimer. Interestingly, it is a very small effect, even for very high fluences ($n_{ex}=5$), which arises from two effects. Firstly, the annihilation is very fast and energy delivery occurs over a finite (and quite long) laser pulse. Secondly, at high fluences absorption competes with SE which means the excitation density never actually reaches $n_{ex}$. Although the second component is visually detectable, bi-exponential re-convolution fits (see below) are not very stable and so are omitted.      

The time-evolution of the system is simulated via a kinetic Monte Carlo algorithm. We define an array of $200$ unconnected LHCII trimers, turn on the simulated laser pulse, and perform Monte Carlo sweeps every $dt = 1 ps$. In a given step each trimer is considered once on average. The occupancy of the trimers and the various rate constants determine the probability of different processes occurring and we consider each process on average once per time-step, $dt$. All processes are assume to be Poissonian and so the probability of process $m$ occurring in the interval $t\rightarrow (t+dt)$ is,
\begin{equation}
p_{m}\left(dt\right)=\nu(dt)e^{-\nu_{m}\left(dt\right)dt}
\label{eq:prob}
\end{equation}
where $\nu_{m}\left(dt\right)$ is the calculated rate. We do not consider the possibility of the same process occurring multiple times in the same trimer in the same time-step and choose $dt = 1$ ps to ensure the likelihood of that occurring is $< 0.1\%$.

We compute the TCSPC trace by binning the times at which 'emissive decays', that is decays via the $k_{\text{decay}}$ channel (though we separately bin all other 'non-emissive' decay processes for book-keeping), until either $J(t)=0$ and $n=0$ for all trimers (i.e. nothing else can happen). A histogram bin width of $\Delta t = 50$ ps is used here, similar to typical experiments. We then reseed the Monte Carlo algorithm, trigger another pulse, and repeat the process until one of the emissive time bins exceeds a threshold count (here chosen to be 10,000 in line with common TCSPC data collection). Due to the definition of the probabilities in Eqn. (\ref{eq:prob}), the traces automatically reproduce the Poissonian noise of TCSPC measurements. 

Lastly, the simulated traces are subject to a \textit{reconvolution fit} in which an exponential decay series,
\begin{equation}\label{eq:decay_components}
F(t)=H(t_{0})\sum_{m}=A_{m}exp\left(-\frac{t-t_{0}}{\tau_{m}}\right)
\end{equation}
is convolved with the IRF, which in our case is the laser pulse. $H(t_{0})$ is simply the Heaviside function used to define the start time, $t_{0}$. Mono-, bi-, and tri-exponential fits are generated and the best is selected on the basis of the variances on the fit parameters and whether the addition of a component visually improves the fit. We then calculate the amplitude-weighted, 
\begin{equation}\label{eq:amp_tau}
\tau_{\text{amp}}=\frac{\sum_{m}A_{m}\tau_{m}}{\sum_{m}A_{m}}
\end{equation}
and intensity-weighted,
\begin{equation}\label{eq:intens_tau}
\tau_{\text{int}}=\frac{\sum_{m}A_{m}\tau_{m}^{2}}{\sum_{m}A_{m}\tau_{m}}
\end{equation}
lifetimes. Throughout we will consider $\tau_{\text{amp}}$ as this is generally the quantity quoted in the literature. The reason for this is that $\tau_{\text{amp}}$ 
is directly proportional to the fluorescence quantum yield (and inversely proportional to the overall extent of quenching) whereas $\tau_{\text{int}}$ is not~\cite{van_Oort_2011}. 

\subsection{Extension of the model to aggregated LHCII (without quenchers)
}

\begin{figure}[ht]
\centering
\includegraphics[width=15cm]{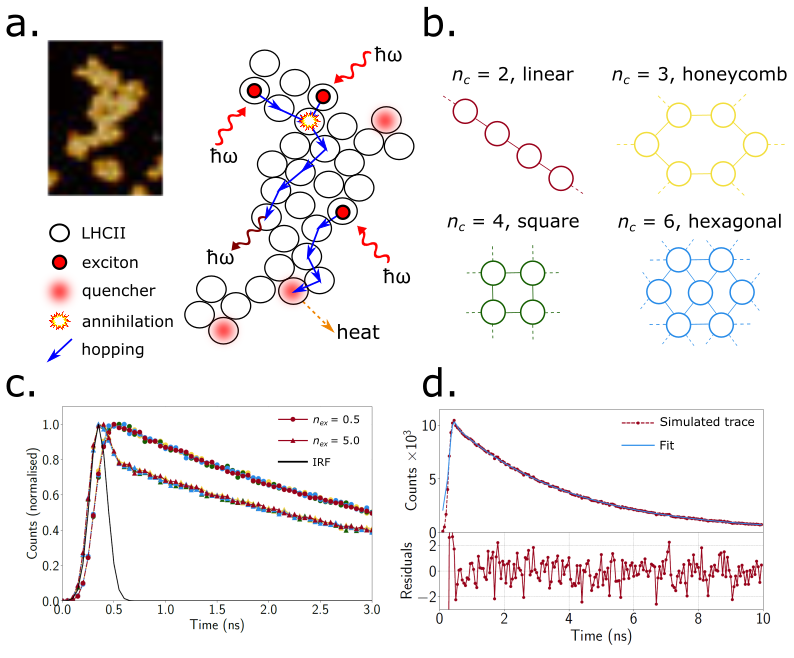}
\caption{\textbf{a.} An AFM image of a pseudo-2D LHCII aggregate on a mica surface~\cite{adams_correlated_2018} (inset), rendered into a discrete coarse-grained model of photon absorption, energy transfer, annihilation, quenching and fluorescence. Each LHCII trimer is treated as a thermodynamic system (see Fig. \ref{fig1}). \textbf{b.} The different lattice geometries considered in our aggregate models. \textbf{c.} The simulated TCSPC kinetics of the different lattice geometries (colour coded as in b.) in the absence ($n_ex=0.5$) and presence ($n_{ex}=5$) of annihilation. The kinetics for different geometries are hard to discern as they are essentially identical. \textbf{d.} An example of a reconvolution fit for a hexagonal lattice and $n_{ex}=2$, shown with the weighted residuals.}
\label{fig2}
\end{figure}

Fig. \ref{fig2} \text{a.} shows a typical laminar aggregate of LHCII, deposited on mica, as visualized by AFM~\cite{adams_correlated_2018}. By simply packing circles of the approximate diameter of an LHCII timer ($~5$ nm) into the area we can construct a very approximate map. It shows that aggregates are small (10-100 trimers) and heterogeneous in terms of coordination number, $n_{c}$, of each trimer. Instead of considering specific aggregates we consider ideal lattices of 100 trimers with $n_c=2$, $3$, $4$, and $6$ (see Fig. \ref{fig2} \textbf{b.}). In the absence of any quenchers the coupled equations of motion of the aggregate become,
\begin{equation}
\begin{split}
\frac{d}{dt}n_{i} &= \sigma J \left(\tilde{N} - n_{i}\right) - \sigma_{\text{SE}}J(t)n_{i} - k_{\text{decay}} n_{i} - \gamma n_{i} \left(n_{i} - 1\right)\\ &- \sum_{j=1}^{n_{c}}{\left(K_{n_{i},n_{j}}^{n_{i}-1,n_{j}+1}n_{i}-K_{n_{i},n_{j}}^{n_{i}+1,n_{j}-1}n_{j}\right)}
\label{eq:dndt_agg_no_quench}
\end{split}
\end{equation}
where $K_{n_{i},n_{j}}^{n_{i}-1,n_{j}+1}$ and $K_{n_{i},n_{j}}^{n_{i}+1,n_{j}-1}$ are the rate constants for the transfer of an excitation from trimer $i$ to neighbouring trimer $j$ and vice versa. They depend on the initial occupancies of the two trimers due to an \emph{entropic repulsion} effect in which the transfer of an excitation to an already crowded trimer is thermodynamically penalized. We define, 
\begin{equation}\label{eq:rate_entropy}
K_{n_{i},n_{j}}^{n_{i}-1,n_{j}+1}=
\begin{cases}
\tau_{hop}^{-1} \quad\text{for} \quad\Delta S_{n_{i},n_{j}}^{n_{i}-1,n_{j}+1}\leq 0 \\
\\
\tau_{hop}^{-1}exp\left(\frac{1}{k_{B}}\Delta S_{n_{i},n_{j}}^{n_{i}-1,n_{j}+1}\right) \quad \text{for}\quad \Delta S_{n_{i},n_{j}}^{n_{i}-1,n_{j}+1} > 0\\
\end{cases}
\end{equation}
where $\tau_{hop}$ is a phenomenological hopping time. Values of $5< tau_{hop} <25 $ ps have been variously assumed in course-grained models of the PSII antenna~\cite{Valkunas2009_CG} and we use $\tau_{hop}=25$ ps, since variations in this range have almost no effect on the simulated TCSPC kinetics other than a tiny re-scaling of the mean lifetime. $\Delta S_{n_{i},n_{j}}^{n_{i}-1,n_{j}+1}$ is the entropy change of for the transfer of the excitation,
\begin{equation}\label{eq:delta_S}
\Delta S_{n_{i},n_{j}}^{n_{i}-1,n_{j}+1}=k_{B} \text{ln}\left[\frac{n_{i}\left(\tilde{N}-n_{j}\right)}{\left(n_{j}+1\right)\left(\tilde{N}-\left(n_{i}-1\right)\right)}\right]
\end{equation}
where this reduces to $K_{1,0}^{0,1}=K_{1,0}^{0,1}\equiv \tau_{hop}^{-1}$ in the single excitation limit. A derivation of Eqn. (\ref{eq:delta_S}) is included in the Supplementary Material. 

Fig. \ref{fig2} \textbf{c.} shows the (lack of) dependence on the TCSPC kinetics on aggregate geometry for low ($n_{ex}=0.5$) and very high ($n_{ex}=5$) fluences. We see that the effect of annihilation is enhanced with respect to the isolated trimers, since the excitations can now hop around the aggregate and encounter each other. Fig. \ref{fig2} \textbf{d.} shows a typical re-convolution fit of these kinetics (shown is $n_{ex}$, the onset of annihilation) including the weighted residuals. From $0 < n_{ex} <2$ the trace is mono-exponential with $\tau_{\text{amp}}=3.6$ ns (by construction). For $n_{ex}\geq 2$ a second component, $\tau_{2}=40 - 140$ ps, is needed and $\tau_{\text{amp}}=1-2$ ns. This is shown in Fig. S1 of the Supplementary material. An excellent quality of fit and low residuals as shown in this example were achieved for all simulations.

\subsection{Extension of the model to include quenching mechanisms}

\begin{figure}[ht]
\centering
\includegraphics[width=10cm]{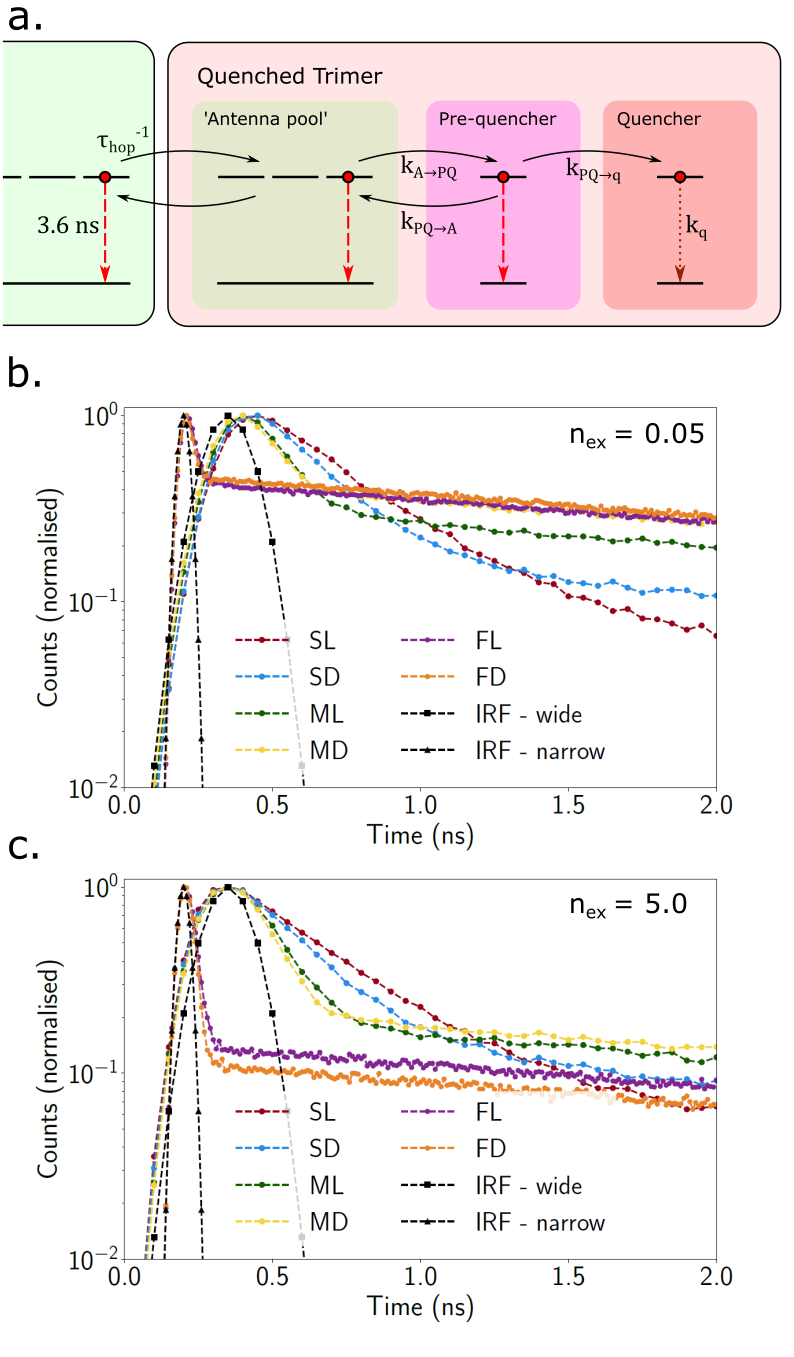}
\caption{\textbf{a.} A schematic of a the generalized thermodynamic model of a quenched trimer (next to an unquenched trimer to the left). The LHCII trimer consists of a reduced number of non-quencher states known as the 'antenna pool'. They can exchange energy with a 'pre-quencher' (or states), which irreversibly transfers energy to the actual 'quencher'. The quencher state dissiaptes energy on timescale of $\Gamma=10$ ps. \textbf{b.} The simulated TCSPC traces for the various quenching models outlined in Table (\ref{tab:params}) at $n_{ex}=0.05$. We have adjusted the quencher density, $\rho_{q}$, in each case to achieve $\tau_{amp}\sim 500$ ps (fits are not shown for clarity). For the slow (SL, SD) and moderate (ML, MD) models a pulse-width of 200 ps (wide IRF) and bin width of $\Delta t = 50$ ps was used. For FL and FD well-behaved fits were only obtained for a $50$ ps pulse, as discussed in main text and shown in Fig. S2 of the Supplementary Material. \textbf{c.} The same as in b. but at high excitation density, $n_{ex}=5$} 
\label{fig3}
\end{figure}

Experimentally, aggregation of LHCII is always accompanied by excitation quenching ($\tau_{\text{amp}}< 2$ ns) and it is assumed that this is due to a sub-population of trimers that are in some quenched state. Rather than address particular published quenching models we adopt a generalized kinetic scheme which is sketched in Fig. \ref{fig3} \textbf{a.}. We assume that there is a quencher with a very short lifetime, $\Gamma$, which we assume is $\sim 10$ ps throughout. This is roughly the excitation lifetime of the various carotenoids proposed as quenchers~\cite{ruban_identification_2007,liguori_different_2017} or the decay time of the Chl-carotenoid CT states in other models~\cite{lapillo_energy_2020} and measurements on LHCII aggregates imply that the quenching state is very short-lived~\cite{chmeliov_nature_2016}. Next, we have a \textit{pre-quencher} which is a sub-set, $\tilde{N}^{pq}$, of the antenna states that the quencher can directly access. Transfer of energy from the pre-quencher to the quencher is assumed to be irreversible, since the quencher functions as an excitation trap. Lastly, we have the \textit{antenna pool}, the rest of the antenna states, $\tilde{N}^{a}-\tilde{N}-\tilde{N}^{pq}$, left unchanged by the formation of the quencher but are coupled to it via the pre-quencher. Energy transfer between the quencher and pre-quencher state is assumed to be bi-directional, depending largely on the balance of  $\tilde{N}^{a}$ and $\tilde{N}^{pq}$. The equations of motion are,
\begin{equation}
\begin{split}
\frac{d}{dt}n_{i}^{a} &= \sigma J \left(\tilde{N}^{a} - n_{i}^{a}\right) - \sigma_{\text{SE}}J(t)n_{i}^{a} - k_{\text{decay}} n_{i}^{a} - \gamma n_{i}^{A} \left(n_{i}^{a} - 1\right)\\ &- \sum_{j\neq i}{\left(K_{n_{i}^{a},n_{j}^{a}}^{n_{i}^{a}-1,n_{j}^{a}+1}n_{i}^{a}-K_{n_{i}^{a},n_{j}^{a}}^{n_{i}^{a}+1,n_{j}^{a}-1}n_{j}^{a}\right)}-k_{a, pq}n_{i}^{a}+k_{pq,a}n_{i}^{pq}
\label{eq:dndt_a}
\end{split}
\end{equation}

\begin{equation}\label{eq:dndt_pq}
\begin{split}
\frac{d}{dt}n_{i}^{pq} &=\sigma J \left(\tilde{N}^{pq} - n_{i}^{pq}\right) - \sigma_{\text{SE}}J(t)n_{i}^{pq} - k_{\text{decay}} n_{i}^{pq} - \gamma n_{i}^{pq} \left(n_{i}^{pq} - 1\right)\\ &-\left(k_{pq,a}+k_{pq,q}\right)n_{i}^{pq}+k_{a,pq}n_{i}^{a}
\end{split}
\end{equation}

\begin{equation}\label{eq:dndt_q}
\frac{d}{dt}n_{i}^{q} =k_{pq,q}n_{i}^{pq}-\frac{1}{\Gamma}n_{i}^{q}
\end{equation}
where $n_{i}^{a}(t)\leq \tilde{N}_{i}^{a}$, $n_{i}^{pq}(t)\leq\tilde{N}^{pq}$, and $n_{i}(t)\leq 1$ are excitation occupations of the antenna pool, pre-quencher and quencher respectively, $k_{a,pq}$ is rate constant for energy transfer \textit{from} the antenna pool \textit{to} the pre-quencher ($k_{pq,a}$ is the reverse), and $k_{pq,q}$ is the transfer rate \textit{from} the pre-quencher to the quencher. 

There is enough flexibility in the model to qualitatively represent most proposed quenching mechanisms, though here we are focusing on those mediated by a carotenid-like (short-lived, optically-forbidden) state. We then consider \textit{fast}, \textit{moderate}, and \textit{slow} quenchers (as compared to $\tau_{\text{hop}}$) as $k_{pq,q}^{-1} = 1$, $25$, and $100$ ps respectively. In a fast quencher an excitation is more likely to be quenched than it is to hop to a neighbouring LHCII trimer and vice versa in a slow quencher. The $<1 \text{ps}$ energy transfer from Chl \textit{a} to lutein observed in the 2D electronic spectra of LHCII in native membranes~\cite{son_observation_2020}, or the Chl-carotenoid excitonic quenching mechanism based on two-photon excitation spectroscopy~\cite{bode_regulation_2009},  would be examples of fast quenchers. By contrast, quenching by incoherent Chl-lutein EET ($\sim 10-50$ ps)~\cite{ruban_identification_2007,fox_carotenoid_2017} or by the formation of Chl-lutein CT states~\cite{ahn_architecture_2008,cupellini_charge_2020}, would be a moderate to slow mechanisms. If we assume the antenna pool and pre-quencher are iso-energetic then $k_{a,pq}/k_{pq,a}$ is dictated the relative sizes of the two domains (essentially entropy)
\begin{equation}\label{eq:k_apq}
\frac{k_{a,pq}}{k_{pq,a}}=exp\left(\frac{1}{k_{B}}\Delta S_{n^{a},n^{pq}}^{n^{a}-1,n^{pq}+1}\right)
\end{equation}
If we assume that the base rate of energy relaxation amongst the Chls in trimer is $\tilde{k}^{-1}\sim 1$ ps then we can define two limiting cases. An \textit{entropically-limited} or \textit{local} quencher,
\begin{equation}\label{eq:k_local1}
k_{a,pq}=\frac{1}{\tilde{N}-1}\tilde{k}
\end{equation}
\begin{equation}\label{eq:k_local2}
k_{pq,a}=\tilde{k}
\end{equation}
is one in which the quencher is connected only to a single Chl state which results in an entropic 'bottleneck' between the antenna Chls and the quencher. Models that propose quenching by the Chl \textit{a}612-Lutein620 heterodimer~\cite{ruban_identification_2007,balevicius_fine_2017,son_observation_2020} are examples of local quenchers. We also have \textit{delocalised} quenchers in which the antenna-pool and the pre-quencher are comparable in size,
\begin{equation}\label{eq:k_delocal}
k_{a,pq}=k_{pq,a}=\tilde{k}
\end{equation}
Mechanisms in which the carotenoid is coupled to several Chls ~\cite{lapillo_energy_2020} are delocalised. Table 1 summarises the different parameter schemes. 

To implement quenching in our TCSPC simulation we  assume that quenching in our aggregate arises from a random fraction, $\rho_{q}=N_{q}/N_{LHCII}$ of the quenched trimers. This random distribution is reinitialized before every excitation pulse ensuring we average over a large number of possible configurations. 

\begin{table}[htb]
    \centering
    \begin{tabular}{|c|c|c|c|}
    \hline
         Quenching Model & $k_{a,pq}^{-1}$ (ps) & $k_{pq,a}^{-1}$ (ps) & $ k_{pq,q}^{-1} (\text{ps}) $  \\
         \hline
         Slow Localised (SL) & 5 & 1 & 100 \\
         Slow Delocalised (SD) & 1 & 1 & 100 \\
         Medium Localised (ML) & 5 & 1 & 25 \\
         Medium Delocalised (MD) & 1 & 1 & 25 \\
         Fast Localised (FL) & 5 & 1 & 1 \\
         Fast Delocalised (FD) & 1 & 1 & 1 \\
         \hline
    \end{tabular}
    \caption{Relevant parameters for the various quenching models used in our LHCII aggregate simulations. In all cases $\Gamma=10$ ps.}
    \label{tab:params}
\end{table}

\section{Results}

\subsection{Simulated TCSPC kinetics of different quenching models}\label{subsec:kinetics}

\begin{figure}[ht]
\centering
\includegraphics[width=11cm]{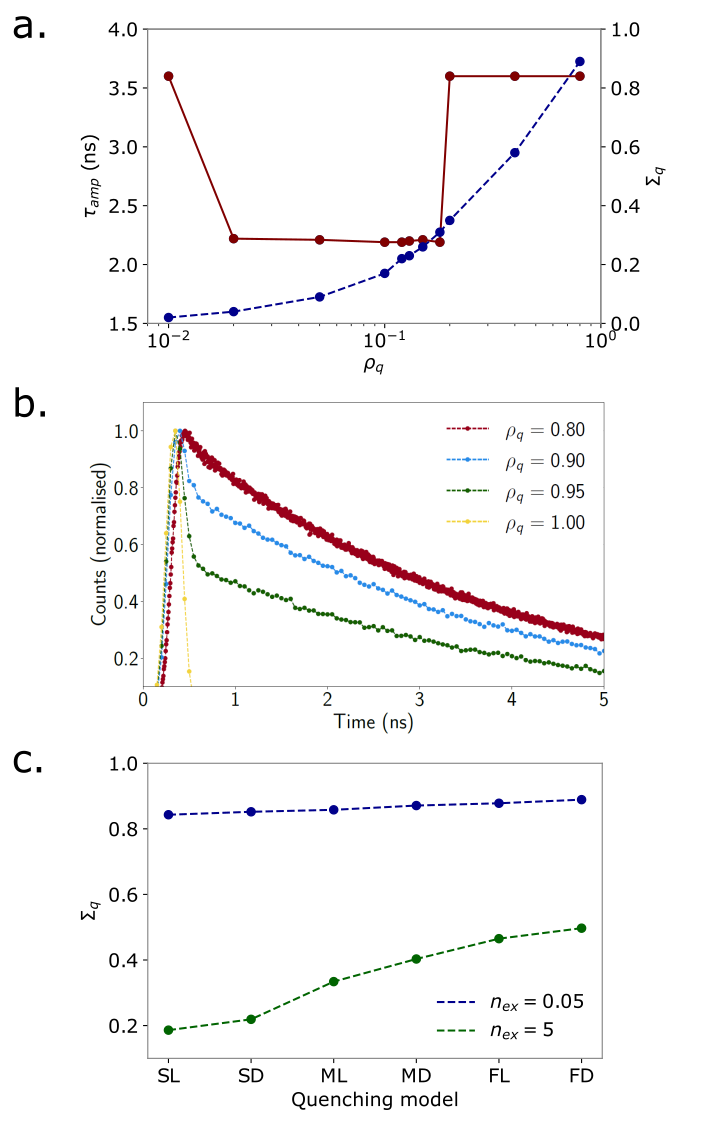}
\caption{\textbf{a.} A plot of $\tau_{\text{amp}}$ as a function of $\rho_{q}$ for an LHCII aggregate with FD quenchers and an assumed pulse-width (IRF) of $\text{FWHM}=200$ ps (red,left vertical axis). For all data points the decay kinetics are purely mono-exponential. Also shown (blue, right vertical axis) is the total fraction of all excitation that are eventually quenched, $\Sigma_{q}$. We see that though $\Sigma_{q}$ increases monotonically with $\rho_{q}$, the 'detected' lifetime is discontinuous with respect to $\rho_{q}$ and never drops below $\tau_{\text{amp}} \sim 2.2$ ns. \textbf{b.} The simulated traces for $0.8\leq \rho_{q} \leq 1.0$. The re-convolution fits were very poor and so it is difficult to assign a $\tau_{\text{amp}}$ in this range. Note the sharp transition from mono-exponential, to sharply multi-exponential, to the pure IRF. \textbf{c.} $\Sigma_{q}$ for the various quenching models. The blue plot corresponds to annihilation-free conditions ($n_{ex}=0.05$) with $\rho_{q}$ chosen to yield $\tau_{\text{amp}}\sim 500$ ps (see Table \ref{tab:fit_no_annihilation}). The green plot corresponds to the same values of $\rho_{q}$ but in conditions of significant excitation annihilation ($n_{ex}= 5$, see Table \ref{tab:fit_annihilation}).} \label{fig4}
\end{figure}

Initially we considered quenching in annihilation-free conditions, choosing $n_{\text{ex}}=0.05$ for an aggregate of $200$ LHCII trimers, meaning that each pulse delivers an average of $10$ excitations. We then took each quenching model listed in Table (\ref{tab:params}) and adjusted the fraction of quenched LHCII, $\rho_{q}$, until we (approximately) obtained a representative quenched lifetime of $\tau_{\text{amp}} \sim 500$ ps (although $200-1200$ ps is generally considered 'quenched'~\cite{pascal_molecular_2005}). Getting an exact match between models was quite difficult as $\tau_{\text{amp}}$ is as sensitive to fit quality as it is to the underlying excitation dynamics. Regardless, it is only qualitative differences in the kinetics that we are interested in as these are probably all that are observable experimentally. The traces are shown in Fig. \ref{fig3} \textbf{b.} (on a logarithmic scale) and the quencher densities and re-convolution fit parameters are listed in Table \ref{tab:fit_no_annihilation}.  

For SL and SD a quencher concentration of $\rho_{q}=0.9$ and $0.85$ respectively are required to reduce the lifetime to $\tau_{\text{amp}}\sim 380 - 530$ ps. Since the kinetics are marginally bi-exponential (particularly SL) we listed both mono- and bi-exponential fits Table \ref{tab:fit_no_annihilation}, with the traces and fits plotted in Figs. S3 and S4 of the Supplementary Material. The bi-exponential fits are visually closer and yield $\tau_{\text{amp}}\sim 530$ and $521$ ps for SL and SD respectively. However, the slow component, $\tau_{1}\sim 4.3 - 5.7$ ns, is significantly slower than the free decay of an unquenched LHCII trimer ($k_{\text{decay}}^{-1} = 3.6$ ns), probably indicating that neither mono- nor bi-exponential fits are perfect. The fast component is $\tau_{2}\sim 264 - 383$ ps, indicating that the 'slow' mechanisms can significantly quench an isolated LHCII trimer (as previously noted~\cite{balevicius_fine_2017}).

For ML and MD a quencher density of $\rho_{q}=0.80$ yields $\tau_{\text{amp}}\sim 568$ and $584$ ps respectively, and kinetics that are obviously bi-exponential (Fig. \ref{fig3} \textbf{b.}). The slow component, $\tau_{1} = 3.7-3.9$ ns is essentially just free decay of unquenched LHCII while the fast component, $\tau_{2} = 167-264$ ps, is comparable to the timescale of the IRF. Interestingly, despite the ML and MD quenchers being much faster than SL and SD ($k_{pq,q}^{-1}=25$ ps compared to $100$ ps) a high density of quenchers is still needed to quench the whole aggregate. This is due to the fact that, since $\tau_{\text{hop}} = k_{pq,q} = 25$ ps, an excitation still has a fairly high probability of evading a quencher by hopping to a neighbouring LHCII trimer.    

The behaviour of FL and FD is complicated by the fact that the kinetics of the quencher are significantly faster than the $200$ ps pulse-width (IRF) initially used in our model, with the decay kinetics having a seemingly counter-intuitive dependence on $\rho_{q}$. In Fig. \ref{fig4} \textbf{a.} (red line) we show the dependence of $\tau_{\text{amp}}$ on $\rho_{q}$ in the range $0.01\leq\rho_{q}\leq 0.8$, for the FD mechanism (the behaviour of FL is identical). The blue line shows the ratio of total excitations quenched, $\Sigma_{q}$, to the total excitations that decay emissively, $\Sigma_{\gamma}$. For very low quencher density ($\rho_{q}\sim 0.01$) the aggregate is essentially unquenched and $\tau_{\text{amp}}\approx k_{\text{decay}}^{-1}$. For $0.02\leq\rho_{q}\leq 0.2$, $\tau_{\text{amp}}$ drops to and plateaus at $\sim 2.2$ ns, indicating a small degree of overall quenching. At $\rho_{q}\sim 0.2$, $\tau_{\text{amp}}$ discontinuously jumps to $3.6 ns$ and remains at this level up to $\rho_{q}\sim 0.8$. Despite the fact that the amount of excitation quenching increases significantly (blue line, Fig. \ref{fig4} \textbf{a.}), it is not 'detected' in the emission histogram. Finally, in the region $0.8\leq\rho_{q}\leq 1.0$ the kinetics become sharply multi-exponential and the reconvolution fits become very unreliable. The interpretation of this is actually straightforward: At low quencher densities there is simply very little quenching. Although FL and FD are labelled 'fast' they are not irreversible traps and therefore excitations have a reasonable chance of evading the quencher. The level of quenching increases sharply at some critical point around $\rho_{q}\sim 0.2$ though it is effectively masked by the wide IRF. Excitations are delivered to the aggregate over $200$ ps but are typically quenched within the first detection bin ($t<50$ ps). All that is detected is the small fraction of excitations that evade the traps and undergo free decay. Essentially, the kinetics are bi-exponential, with the short component completely masked by the IRF. At $\rho_{q}=1.0$, all LHCII trimers are equivalent and deeply quenched and the TCSPC signal is simply the IRF. In the range $0.8\leq\rho_{q}< 1.0$ there is a sharp transition between these two regimes. 

The only way to clearly resolve the FL and FD mechanisms is to reduce the pulse width to $\text{FWHM}=50$ ps and the bin width to $\Delta t=10$ ps (reducing the bin width alone does not significantly alter the kinetics as shown in Fig. S2 of the Supplementary Material). This higher resolution is still experimentally feasible but requires high performance instrumentation. At a quencher density of $\rho_{q}\sim0.8$ the FL and FD mechanisms result in sharply bi-exponential TCSPC kinetics with $\tau_{\text{amp}} \sim 652$ and $538$ ps respectively. The slow component is, as with ML and MD, $\tau_{1}\sim k_{\text{decay}}^{-1} = 3.6$ ns, while the fast component is $\tau_{2}\sim 34$ and $25$ ps. The fits are shown in Fig. S6 of the supplementary material and it is apparent that longer lifetime of the FL model ($\tau_{\text{amp}} \sim 652$ ns) is the result of poor fit quality.      

Lastly, to confirm that we were truly comparing comparable quenching regimes across the different models, we compared $\Sigma_{q}$ at $\tau_{\text{amp}}\sim{500}$ ps (see Fig. \ref{fig4} \textbf{c.}). In all cases there is significant quenching, with $0.84\leq\Sigma_{q}\leq0.89$ for all models. This shows that all mechanism have the some functional consequences, quenching 80-90\% of the energy absorbed by the aggregate, the experimental signatures are very different.

\begin{table}[htb]
    \centering
    \begin{tabularx}{0.95\textwidth}{|c|c|c|c|c|c|c|}
        \hline
        \multicolumn{7}{|c|}{$n_{ex}=0.05$} \\
        \hline
         Model & $\rho_q$ & $\tau_{\text{amp}}$ (ps) & $A_1$ & $ \tau_1 $ (ps) & $A_2$ & $\tau_2$ (ps) \\
         \hline
         SL$_{\text{mono}}$ & 0.9 & $434 \pm 5$ & $1.683 \pm 0.006$ & $434 \pm 5$ & --- & --- \\
         SL$_{\text{bi}}$ & 0.90 & $ 530 \pm 35 $ & $ 0.049 \pm 0.002 $ & $ 5750 \pm 1132 $ & $ 1.729 \pm 0.005 $ & $ 383 \pm 5 $\\
         SD$_{\text{mono}}$ & 0.85 & $ 386 \pm 11 $ & $ 1.765 \pm 0.016 $ & $ 386 \pm 11 $ &  ---  & ---  \\
         SD$_{\text{bi}}$ & 0.85 & $ 521 \pm 16 $ & $ 0.137 \pm 0.000 $ & $ 4298 \pm 219 $ & $ 2.008 \pm 0.007 $ & $ 264 \pm 3 $ \\
         ML & 0.80 & $ 568 \pm 18 $ & $ 0.284 \pm 0.004 $ & $ 3932 \pm 135 $ & $ 2.382 \pm 0.024 $ & $ 167 \pm 4 $ \\
         MD & 0.80 & $ 584 \pm 11 $ & $ 0.403 \pm 0.003 $ & $ 3774 \pm 60 $ & $ 2.806 \pm 0.032 $ & $ 126 \pm 3 $ \\
         FL & 0.80 & $ 652 \pm 4 $ & $ 0.436 \pm 0.001 $ & $ 3645 \pm 14 $ & $ 2.113 \pm 0.012 $ & $ 34 \pm 4 $ \\
         FD & 0.80 & $ 538 \pm 3 $ & $ 0.470 \pm 0.001 $ & $ 3618 \pm 10 $ & $2.824\pm 0.018$ & $25\pm 3$ \\
         \hline
    \end{tabularx}
    \caption{Fit parameters for each quenching model, assuming $n_{ex}=0.05$ and tuning $\rho_{q}$ to achieve an amplitude-weighted lifetime, $\tau_{\text{amp.}}$ of roughly $500$ ps. We show SL and SD parameters for both mono- and bi-exponential fits and note that $\tau_{\text{amp}} \sim 650$ ps for the FL model is due to a poor fit rather than any significant differences in the trace itself.}
    \label{tab:fit_no_annihilation}
\end{table}

\subsection{Different quenching models at high excitation density}

Using the same excitation densities, $\rho_{q}$, as in Table \ref{tab:fit_no_annihilation}, the excitation density was increased to $n_{ex}=5$ to introduce annihilation (traces shown in Fig. \ref{fig3} \textbf{c.}). Visually, the SL and SD models do not appear to have changed, the ML and MD traces seem to decay a little faster and the FL and FD traces are noticeably altered. This is confirmed by the reconvolution fits, for which the parameters are listed in Table (\ref{tab:fit_annihilation}) and the plots shown in Figs. S7 - S10 in the Supplementary Material. The SL and SD mechanisms remain essentially unchanged by the presence of annihilation, though there is some slight difference for SD. The reason for this is the large difference in timescales between SL/SD quenching and annihilation, coupled with the fairly wide excitation pulse. Annihilation is fast and with relatively slow excitation delivery, is largely over by the end of the pulse. SD/SL quenching occurs later and so to two processes do not compete in a way that is detectable in the signal. ML/MD and particularly FL/FD do compete with annihilation and so their kinetics are visibly dependent on excitation density. This can also be seen in Fig. \ref{fig4} \textbf{c.} which shows the decrease in $\Sigma_{q}$ due to annihilation, which is most significant for the SL and SD quenchers.  

\begin{table}[htb]
    \centering
    \begin{tabularx}{0.95\textwidth}{|c|c|c|c|c|c|c|}
        \hline
        \multicolumn{7}{|c|}{$n_{ex}=5.0$}\\
        \hline
         Model & $\rho_q$ & $\tau_{\text{amp}}$ (ps) & $A_1$ & $\tau_1 $ (ps) & $A_2$ & $\tau_2$ (ps) \\
         \hline
         SL$_{\text{mono}}$ & 0.9 & $433 \pm 5$ & $1.664 \pm 0.008$ & $433 \pm 5$ & --- & --- \\
         SL$_{\text{bi}}$ & 0.90 & $524 \pm 36$ & $0.047 \pm 0.003$ & $ 5723 \pm 1186 $ & $ 1.714 \pm 0.009 $ & $ 381 \pm 5 $ \\ 
         SD$_{\text{mono}}$ & 0.85 & $377 \pm 8$ & $1.759\pm 0.008$ & $377\pm 8$ & --- & ---\\
         SD$_{\text{bi}}$ & 0.85 & $516\pm 36$ & $1.900\pm 0.012$ & $5186\pm702$ & $0.089\pm 0.004$ & $297\pm 6$ \\
         ML & 0.80 & $449\pm 31$ & $0.143\pm 0.004$ & $4699\pm 474$ & $2.410\pm 0.022$ & $197\pm 6$\\
         MD & 0.80 & $428\pm 28$ & $0.169\pm 0.005$ & $4481\pm 402$ & $2.674\pm 0.028$ & $172\pm 6$\\
         FL & 0.80 & $192\pm 9$ & $0.086\pm 0.002$ & $4275\pm 232$ & $2.407\pm 0.012$ & $46\pm 1$ \\
         FD & 0.80 & $193\pm 8$ & $0.095\pm 0.002$ & $4196\pm 211$ & $2.520\pm 0.014$ & $43\pm 1$\\
         \hline
    \end{tabularx}
    \caption{Fit parameters for the same models (and model parameters) but for $n_{ex}=5.0$.}\label{tab:fit_annihilation}
\end{table}

\subsection{Measured fluorescence kinetics of LHCII aggregates}\label{subsec:experiment}

\begin{figure}[ht]
\centering
\includegraphics[width=11cm]{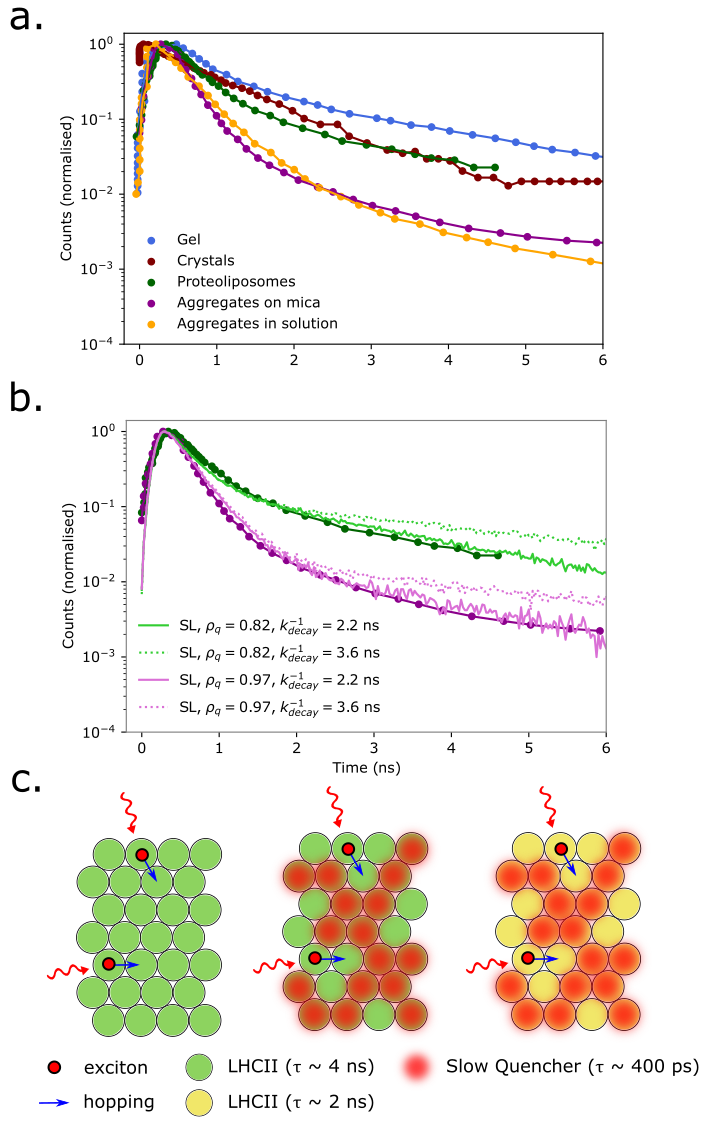}
\caption{\textbf{a.} A selection of TCSPC measurements on quenched LHCII in a variety of experimental conditions. These include LHCII crystal~\cite{pascal_molecular_2005}, LHCII immobilized in gel~\cite{ilioaia_induction_2008}, and LHCII aggregates in proteoliposomes~\cite{natali_light-harvesting_2016}, solution~\cite{johnson_photoprotective_2009}, and on a mica surface~\cite{adams_correlated_2018}. Data was obtained by the WebPlotDigitizer service~\cite{Rohatgi2022}. \textbf{b.} The measured TCSPC trace for LHCII aggregates in proteoliposomes~\cite{natali_light-harvesting_2016} (green dots) and on mica~\cite{adams_correlated_2018} (purple dots) shown with various simulated traces for an aggregate with SL quenchers. The only differences between the two traces appears to be the value of $\rho_{q}$. The dotted lines are the best fits obtainable with the constraint $k_{\text{amp}}^{-1}=3.6$ ns. The fits are significantly improved by relaxing this constraint and setting $k_{\text{amp}}^{-1}=3.6$ ns (solid lines). \textbf{c.} A schematic of different aggregate scenarios. On the \textbf{\textit{left}} there is the hypothetical aggregate in the absence of quenching. Each site is identical with a free excitation decay time of $k_{\text{decay}}^{-1}\sim 4$ ns. In the \textbf{\textit{middle}} is a conventional picture of quenching in which a sub-population of trimers adopt a quenching configuration. On the \textbf{\textit{right}} is the scenario implied by experimental measurements in which, in addition to the quenchers, there is global decrease in the free Chl lifetime, $k_{decay}^{-1}$} 
\label{fig5}
\end{figure}

TCSPC measurements on quenched LHCII aggregates have been reported in a variety of experimental conditions. Here we compare (annihilation-free) traces from LHCII crystals~\cite{pascal_molecular_2005}, LHCII trimers in gel~\cite{ilioaia_induction_2008}, in proteoliposomes~\cite{natali_light-harvesting_2016}, aggregates in solution~\cite{johnson_photoprotective_2009}, and laminar aggregates on mica~\cite{adams_correlated_2018}. The published traces were digitised using the WebPlotDigitizer service~\cite{Rohatgi2022} and are plotted (normalized) in Fig. \ref{fig5} \textbf{a.} Visually, they seem to resemble our SD and SL model traces, lacking the sharply and obviously bi-exponential kinetics of the faster quenching models. Still, the kinetics are clearly heterogeneous and they are generally reported with a bi-exponential fit consisting of a 'slow' ($\sim 2$ ns) and a 'fast' ($\sim 200-600$ ps) component. The fast component matches the $\tau_{2}\sim 300-400$ ps component of our SL and SD models, which reflects migration to and then trapping by a very large distribution of slow quenchers. In all of our models the slow component, $\tau_{1}\sim 3-5$ ns, essentially reflects the free decay of excitations in unquenched trimers. 

Using our simulations we attempted to fit some of these experimental traces. We focused on LHCII in proteolipsomes~\cite{natali_light-harvesting_2016} and LHCII aggregates on a mica surface~\cite{adams_correlated_2018} since these are quasi-2D arrays that should resemble our model. LHCII crystals and aggregates in solution, despite being highly non-native structures, produce very similar TCSPC kinetics to LHCII in proteolipsomes and on mica respectively (but with a narrower IRF). LHCII in gel is equally non-native, with gel-immobilization assumed to induce quenching in the absence of aggregation. The fits are plotted in Fig. \ref{fig5} \textbf{b.} Both experimental traces could be fit to an SL quenching model, assuming $\rho_{q}=0.82$ and $0.97$ respectively. The dotted lines show the best fits we could obtain with the constraint that $k_{\text{decay}}^{-1}=3.6$ ns (taken from fitting LHCII in detergent in Fig. \ref{fig1} \textbf{b.}) and while the short time kinetics are fit very well the longer times are not. The closest visual fits (solid lines in Fig. \ref{fig5} \textbf{b.}) are obtained by setting $k_{\text{decay}}^{-1}=2.2$ ns implying that, in addition to a large sub-population of slow, localized quenchers, there is some global reduction of Chl excited state lifetime. A schematic diagram illustrating this \textit{mixed quenching} scenario is shown in Fig. \ref{fig5} \textbf{c.}

\section{Discussion}

The essential aim of this work was to determine whether one could identify the nature of the NPQ quencher from a relatively 'simple' TCSPC measurement on an LHCII aggregate. It had been previously suggested that this may be the case if one considered excitation densities high enough to induce singlet-singlet annihilation~\cite{valkunas_excitation_2011}. We sought to evolve this work in two ways. Firstly, we aimed to model, as closely as possible, the actual experiment, including finite instrument responses, limited temporal resolution and reconvolution fitting of the resulting signal, all factors that could completely mask any fine kinetic detail of the quenching mechanism. Secondly, various works combining quantum chemical modelling of isolated LHCII trimers with ultra-fast optical spectroscopies have provided us with a selection of plausible (if not definitive) quenching mechanisms, meaning we can focus our model on a narrower parameters space than previously considered. 

\subsection{The TCSPC kinetics of different quenching mechanisms}\label{subsec:different}

In previous work it was argued that two limiting models of quenching, fast trapping and slow quenching (F/S) and vice-versa (S/F), have identical fluorescence kinetics in the absence of annihilation but different dependencies on increasing excitation density~\cite{valkunas_excitation_2011}. While it is difficult to compare directly due to model difference, our results are rather different. Firstly, we see stark differences in the simulated signals between different quenching scenarios, even at very low excitation density. Our SL and SD models, which are closest to the previous S/F model, result in fluorescence kinetics that are almost mono-exponential, but for a small-amplitude, long-time tail. Conversely, the faster quenching mechanisms are sharply bi-exponential. For faster quenching models the difference in the lifetimes of quenched and unquenched LHCII trimers is larger and fewer LHCII trimers need to be quenched to obtain the same overall lifetime, $\tau_{\text{amp}}$, for the aggregate.

When we introduced annihilation the TCSPC kinetics of the SL and SD models barely change, while those of the faster models are associated with a significant drop in $\tau_{\text{amp}}$. This is contrary to previous theoretical study in which slow-trapping (S/F) quenching models were suggested to be far more sensitive to excitation density than fast-trapping (F/S) models ~\cite{valkunas_excitation_2011}. The rational was that fast-trapping quenchers trap excitons before they have a chance to annihilate while slow-trapping quenchers don't. The reason we see the opposite in our more realistic simulations is as follows: 

\begin{itemize}
\item{In a TCSPC measurement excitations are delivered over a time period defined by the pulse width, typically $\sim 50-200$ ps. Since annihilation occurs on the order of ($\gamma^{-1}\sim 20$ ps) the majority of annihilation occurs inside the pulse. Add to this the contribution of stimulated emission and it is quite difficult to accumulate a large \textit{residual} exciton population outside of the pulse (remember $n_{ex}$ is crude estimate of excitation density). The majority of the annihilation is effectively masked by the IRF, unless more expensive/advanced lasers with ultra-narrow temporal width are used.}
\item{The remaining excitons can still diffuse about the aggregate but annihilation is hindered by their mutual \textit{entropic repulsion}, which means they have a tendency to avoid occupying the same timer. This is why annihilation kinetics have little dependence on aggregate topology (Fig. \ref{fig2} \textbf{c.}).}
\item{For SL and SD almost every trimer is a quencher site meaning quenching is by far the most likely dissipation pathway regardless, up to a point, of the residual excitation density.}
\item{For ML and MD a slightly smaller population of quenchers is required to give the same overall level of quenching, meaning the likelihood of two residual excitons meeting each other and annihilating increases slightly. There is therefore a small dependence of $\tau_{amp}$ on excitation density.}
\item{The very large decrease in lifetime for the FL and FD models at high excitation density is primarily the result of the narrower laser pulse ($\sim 50$ ps) needed to resolve the quenching. Excitations are delivered over a shorter timescale, meaning that a higher transient population is achieved and more excitons are present after the pulse. This is also the reason annihilation is so obvious in TA measurements which use sub-picosecond excitation pulses.}
\end{itemize}

This implies that annihilation may not be very useful in discriminating between different quenching mechanisms in real measurements. Different mechanisms should have different fluorescence kinetics even at low excitation densities and the annihilation kinetics are more sensitive to the excitation pulse width than to the quenching mechanism.   

\subsection{The identity of the quencher in NPQ}

Surprisingly, TCSPC measurement, when analysed with our simulated measurements, may yield significant information about the mechanism responsible for quenching in various LHCII assemblies (Fig. \ref{fig5} \textbf{a.} and \textbf{b.}). For LHCII aggregates on mica surfaces~\cite{adams_correlated_2018} and in proteolipsomes~\cite{natali_light-harvesting_2016}, the measured fluorescence kinetics are consistent with a very large sub-population ($\rho_{q}\sim 80-90\%$ or trimers) of 'slow quenchers'. That is some quenching configuration in which energy is delivered slowly ($k_{pq,q}^{-1}\sim 100$ ps) but irreversibly to a trap state with a short lifetime ($\Gamma\sim 10$ ps). Interestingly, the only difference betweenthe two appears to be the quencher density, $\rho_{q}$. This seems in line with very recent time-resolved fluorescence measurements on LHCII in model membranes, which showed that $\rho_{q}$ depends on the lipid-LHCII ratio~\cite{manna_membrane-dependent_2021}. However, whether this slow NPQ trap state is coupled only to particular Chl (SL) or accessible from many (SD) is probably not resolvable with TCSPC, since the fluorescence kinetics for each scenario are \emph{qualitatively} identical. 

Of course we did not explore the entire parameter space of our model. For example, we did not consider a state that a quencher with a fast trapping time ($k_{pq,q}^{-1}\sim 1$ ps) and slow quenching time ($\Gamma > 100 ps$), mainly because this is not suggested by the majority of plausible models. Most models assume the quencher is a carotenoid~\cite{son_observation_2020,lapillo_energy_2020,balevicius_fine_2017,ruban_identification_2007} or some short-lived Chl-carotenoid excitonic~\cite{bode_regulation_2009} or CT state~\cite{park_chlorophyllcarotenoid_2019,cupellini_charge_2020}. Models aside, it was observed in early TA measurements of LHCII aggregates~\cite{ruban_identification_2007} that the quencher did not accumulate a detectable population, implying it was both short-lived and slowly-populated (hence the initial hypothesis of a carotenoid as a quencher). 

Our models do imply that a fast-trapping ($k_{pq,q}^{-1}<10$ ps), fast-quenching ($\Gamma\sim 10$ ps) mechanism would yield very sharply bi-exponential kinetics, something that doesn't appear to be reported despite a wide range of experimental conditions. Of course, one could argue that this behaviour could be masked by a combination of a wide excitation pulse and low detection resolution, but this seems unlikely given the number of independent measurements in vastly different conditions. However, two recent measurements have suggested some form of ultra-fast transfer to a carotenoid or carotenoid-like state is involved in quenching. TA measurements on quenched LHCII immobilized in gel and in low detergent conditions showed the instantaneous (within resolution) appearance of a carotenoid-like signal upon excitation of Chl \textit{a} or Chl \textit{b}~\cite{saccon_protein_2020}. Similarly, 2D measurements of LHCII in lipid nano-disks showed $<500$ fs Chl-carotenoid EET~\cite{son_observation_2020}. While one could argue that the gel immobilization represents a highly non-native set of conditions, it is harder to argue this for lipid nano-disks. 

The most interesting revelation is that at least two quenching mechanisms are needed to reproduce the observed fluorescence kinetics. The reduction of \emph{both} lifetime components has always been a feature of the measured fluorescence kinetics of quenching but it was never clear how this related to the actual internal processes of exciton migration, trapping and decay. The 'second' quenching pathway appears to be rather slow ($\sim 2$ ns) and uniformly distributed within the aggregate. We modelled this as an \textit{ad hoc} reduction in $k_{\text{decay}}^{-1}$, but we are not necessarily proposing that some process radically alters the fundamental excited state lifetime of Chl. We are not excluding this possibility either, given that it is well known that non-planar distortions to tetrapyroles can significantly shorten their fluorescence lifetimes (though generally by enhancing triplet formation)~\cite{Drain_distorted,Chirvony_distorted}. Either way, this shortening of the LHCII lifetime from $\sim 4$ ns to $\sim 2.2$ ns is also observed in the thylakoid membrane in the absence of quenching. It was initially thought to be due to radical pair recombination in the PSII reaction centre ~\cite{Holzwarth_RP,miloslavina2006charge}. However, the same lifetime reduction was observed in plants depleted of PSII due to being grown on the PSII repair inhibitor lincomycin, implying that $\sim 2.2$ ns lifetime comes from some decay process in the antenna~\cite{Belgio_linco}. Our previous energy relaxation simulation of LHCII suggested while simple EET to the carotenoid lutein could not explain the level of quenching seen in NPQ, it may explain the 'background quenching' responsible for the $\sim 2.2$ ns lifetime.       

\subsection{Why focus on Time-Correlated Single Photon Counting?}

High resolution non-linear spectroscopies (TA, 2D, transient grating, etc.) are the state-of-the art technique for unravelling the ultra-fast and branching relaxation processes in isolated photosynthetic complexes, their utility vastly enhanced by an extremely well-developed theoretical framework~\cite{mukamel_spec,vanAmerongen_excitons}. Still, they have yet to unambiguously reveal key functional aspects such as the underlying mechanism of NPQ. This is possibly due to a combination of a lack of fine structural information on the quenched and unquenched states, and the fact that these measurements generally have to be performed in highly non-native conditions. 

To understand the actual functional characteristics of NPQ one must consider its place within a larger network of complexes, the clusters of LHCII that form in the membrane and even the thylakoid grana as a whole. These length- and time-scales are better probed by techniques such as Fluorescence Lifetime Imaging (FLIM) which, when correlated with visualization techniques such as AFM, offer insight into the relationship between membrane organization and function~\cite{adams_correlated_2018}. FLIM is based on TCSPC and direct data analysis is generally limited to reconvolution fitting of the fluorescence decay kinetics. Models of the underlying processes have in the past been highly phenomenological or overlooked the technical limitations of the method. We have shown here that while factors such as pulse width and detection resolution can have a profound impact on the kinetics and the subsequent fits, this is not necessarily a problem, so long as they are carefully factored into any model as they have been here. 

In conclusion, we here present a first look at new method for analysing the \emph{measured} fluorescence kinetics of large networks of photosynthetic complexes. The approach is general and easily extendable to include non-trivial structures and additional decay pathways such as photochemical quenching in the reaction centre. 

\section{Acknowledgements}

All authors acknowledge the support of BBSRC (joint grant BB/T000023/1).

\section{Software details}\label{sec:sim}

We use a mix of Fortran with OpenMPI for the TCSPC simulation and Python for the fitting and analysis. The pRNG used in the Fortran code is \texttt{xoshiro1024} (see \href{https://gcc.gnu.org/onlinedocs/gcc-9.5.0/gfortran/RANDOM_005fNUMBER.html}{here}) and the code itself is available at \url{https://github.com/QMUL-DuffyLab/aggregates}.

\printbibliography

\end{document}